\documentclass[%
 reprint,
 superscriptaddress,
 amsmath,amssymb,
 aps,
]{revtex4-1}

\usepackage{graphicx}
\usepackage{dcolumn}
\usepackage{bm}
\usepackage[utf8]{inputenc}
\begin{document}

\title{Strongly correlated superconductor with polytypic 3D Dirac points}

\author{S. V. Borisenko}
 \affiliation{IFW Dresden, Helmholtzstr. 20, 01069 Dresden, Germany}
\author{V. Bezguba}
 \affiliation{IFW Dresden, Helmholtzstr. 20, 01069 Dresden, Germany}
 \affiliation{Kyiv Academic University, 03142 Kyiv, Ukraine}
\author{A. V. Fedorov}
 \affiliation{IFW Dresden, Helmholtzstr. 20, 01069 Dresden, Germany}
 \affiliation{Helmholtz-Zentrum Berlin f\"ur Materialien und Energie, Albert-Einstein-Str. 15, 12489 Berlin, Germany}
\author{Y. S. Kushnirenko}
 \affiliation{IFW Dresden, Helmholtzstr. 20, 01069 Dresden, Germany}
\author{V. Voroshnin}
 \affiliation{Helmholtz-Zentrum Berlin f\"ur Materialien und Energie, Albert-Einstein-Str. 15, 12489 Berlin, Germany}
\author{M. Sturza}
 \affiliation{IFW Dresden, Helmholtzstr. 20, 01069 Dresden, Germany}
\author{S. Aswartham}
 \affiliation{IFW Dresden, Helmholtzstr. 20, 01069 Dresden, Germany}
\author{A. Yaresko}
 \affiliation{Max-Planck-institute for Solid State Research, Heisenbergstrasse 1, 70569 Stuttgart, Germany}

\begin{abstract}
Topological superconductors \cite{sato2017topological} should be able to provide essential ingredients for quantum computing, but are very challenging to realize. 
Spin-orbit interaction in iron-based superconductors opens the energy gap between the $p$-states of pnictogen and $d$-states of iron very close to the Fermi level, and such $p$-states have been recently experimentally detected \cite{borisenko2016direct}. Density functional theory predicts existence of topological surface states within this gap in FeTe$_{1-x}$Se$_x$ making it an attractive candidate material \cite{wang2015topological}. Here we use synchrotron-based angle-resolved photoemission spectroscopy and band structure calculations to demonstrate that FeTe$_{1-x}$Se$_x$ (x=0.45) is a superconducting 3D Dirac semimetal hosting type-I and type-II Dirac points and that its electronic structure remains topologically trivial. We show that the inverted band gap in FeTe$_{1-x}$Se$_x$ can possibly be realized by further increase of Te content, but strong correlations reduce it to a sub-meV size, making the experimental detection of this gap and corresponding topological surface states very challenging, not to mention exact matching with the Fermi level. On the other hand, the $p-d$ and $d-d$ interactions are responsible for the formation of extremely flat band at the Fermi level pointing to its intimate relation with the mechanism of high-T$_c$ superconductivity in IBS.
\end{abstract}

\pacs{Valid PACS appear here}
\maketitle

\medskip

Electronic structure of iron-based superconductors (IBS) has been studied very intensely during the last decade. The presence of orbital dependent renormalization, blue-red shifts, temperature dependence, nematic and spin-orbit splitting has been firmly established. The observation of the Fe $d_{3z^2-r^2}$ band  hybridized with As $p_z$ in LiFeAs \cite{borisenko2016direct} raised the question, whether its participation in the formation of the electronic structure near the Fermi level in other IBS is significant. The presence of $p_z$-band has been then confirmed in LiFeAs \cite{wang2015topological} and identified in NaFeAs \cite{watson2018three}. Recent ARPES studies suggested that IBS universally host the topological surface states  within the gap opened due to interaction of $p_z$- and $d_{xz/yz}$-bands in the vicinity of the Fermi level \cite{wang2015topological, zhang2018observation, zhang2019multiple}. However, theoretically predicted size of the inverted gap in FeTe$_{1-x}$Se$_x$  is only of the order of 2-10 meV \cite{borisenko2016direct, wang2015topological}. This value has to be divided by the renormalization factor, universally present in all IBS and known to vary between 3 and 20 depending on the orbital in FeTe$_{1-x}$Se$_x$ (e.g. \cite{tamai2010strong, maletz2014unusual, fedorov2016effect, Watson2015a}). Recent observation of topological surface states in FeTe$_{1-x}$Se$_x$ (x=0.45) \cite{zhang2018observation} by 7 eV laser ARPES implies the size of the inverted gap of more than 20 meV, which is in sharp contrast with the predictions of density-functional theory (DFT). It remains unclear which interaction is responsible for such an enormous new energy scale in IBS. To resolve this controversy we have conducted a detailed high-resolution study of nominally the same material using the photons of many different energies to determine its bulk electronic structure. Our results show that all of the observed features are in a qualitative agreement with DFT. We identify all bulk states and track their $k_z$-dispersion, but find no evidence for the surface states. The behavior of bulk states implies the presence of 3D Dirac points of both types and anomalously strong renormalization, while electronic structure of FeTe$_{1-x}$Se$_x$ (x=0.45) remains topologically trivial.

\begin{figure*}
	\centering
		\includegraphics[width=0.95\linewidth]{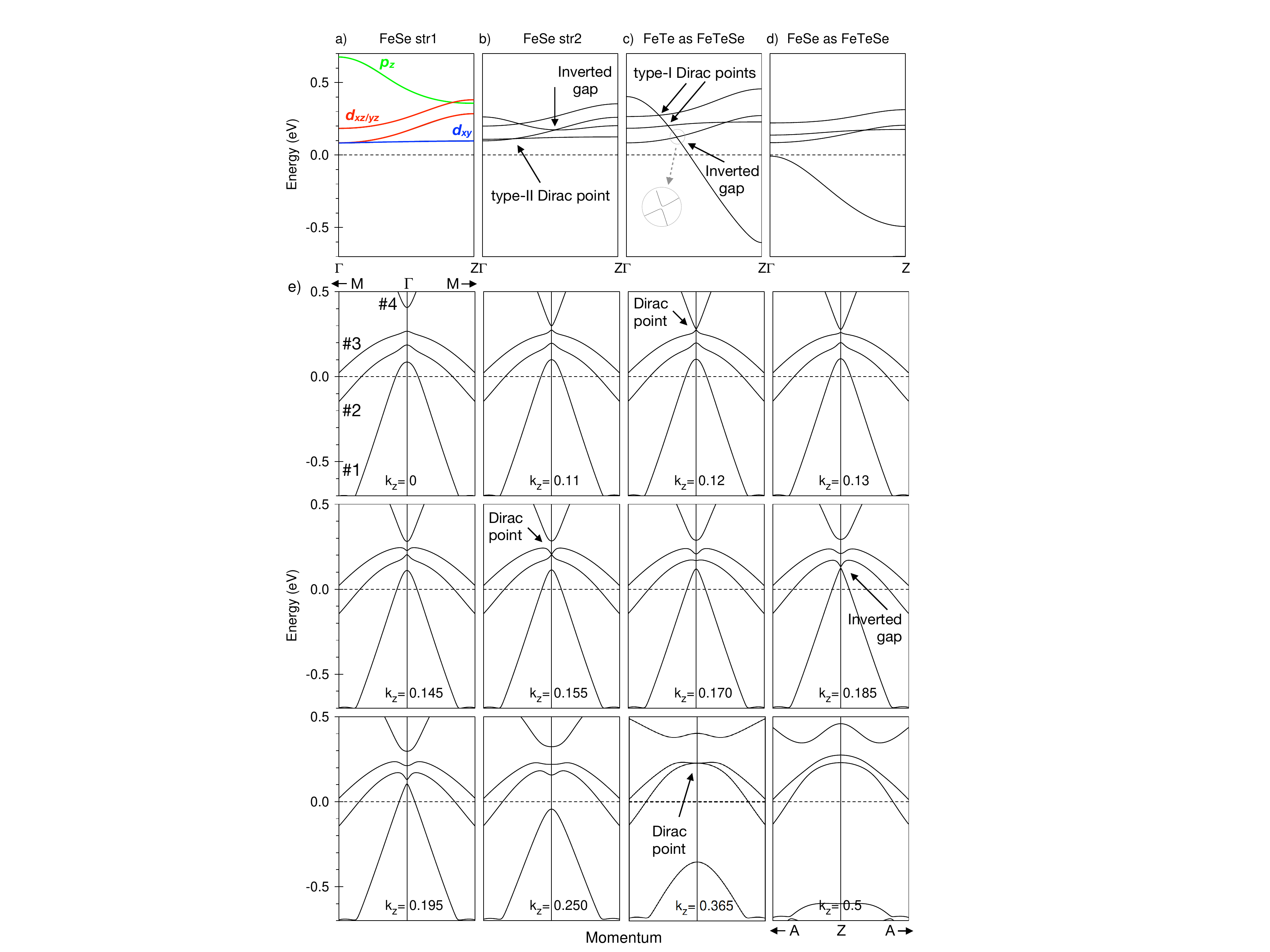}
	\caption{(a-d) Band structure calculations of FeTe$_{1-x}$Se$_x$ along $\Gamma$Z for different crystal structures. (e) Bands in the case from panel (c) along momentum cuts parallel to M$\Gamma$M and AZA for selected values of $k_z$. }
	\label{fig:1}
\end{figure*}

To get an insight into the band structure of a material with a partial substitution, \textit{ab-initio} calculations should take into account unavoidable uncertainties related to the crystal structure. In Fig. \ref{fig:1}(a-d) we present the results of the band structure calculations along the $\Gamma$Z-direction performed for several crystal structures including two refinements for FeSe and one for FeTe$_{1-x}$Se$_x$ with only Te or only Se atoms, respectively \cite{li2009first, bohmer2013lack, tegel2010crystal, khasanov2010iron, mcqueen2009extreme}. There are three dispersions which gain energy with momentum and one which does the opposite, all behaving qualitatively similar in all four cases. Although we label these dispersions traditionally as $d_{xz/yz}$, $d_{xy}$ and $p_z$ and assign the corresponding color (see Fig. \ref{fig:1}a), one has to realize that with the presence of spin-orbit splitting the orbital composition of particular states becomes non-trivial, especially along the high-symmetry directions in the Brillouin zone. Several important conclusions may be drawn from the presented calculations. Fully tellurium-substituted material results in approximately twice stronger spin-orbit splitting of the $d_{xz/yz}$-states (shown in red in Fig. \ref{fig:1}a), which remains constant along $\Gamma$Z in all cases. Slightly varying energetics of $d_{xy}$-states with respect to $d_{xz/yz}$-states defines the momentum location of the 3D Dirac point of the second type, marked in Fig. \ref{fig:1}b \cite{soluyanov2015type, borisenko2019time}, which is universally present in all four cases. The most of the diversity in band structures shown in Fig. \ref{fig:1}(a-d) is brought by the $p_z$-dispersion of the chalcogen. Depending on the structure, this band can cross none, one, two or all three iron-derived $d$-dispersions. This is the well-known result of sensitivity of the $p$-states to the chalcogen height \cite{wu2016topological}, and thus a thorough experimental study is needed to find out which of the sketched scenarios is realized in FeTe$_{0.55}$Se$_{0.45}$. All crossings in Fig. \ref{fig:1}(a-d) are 3D Dirac points protected by symmetry except the one where $p_z$-dispersion reaches the lower one of the $d_{xz/yz}$-dispersions (Fig. \ref{fig:1}b, c). It is in this situation that the tiny energy gap opens and band inversion occurs, i.e. the system becomes topologically nontrivial, causing the appearance of topological surface states \cite{wang2015topological}.

Before we turn to the analysis of ARPES data, we consider in more details the most complicated scenario shown in Fig. \ref{fig:1}c, when the maximal number of band crossings occur and the inverted band gap opens. In a typical ARPES experiment aimed to probe electronic structure in 3D momentum space one measures the energy-momentum intensity distribution as a function of photon energy. We plot the band structure along such momentum cuts in Fig. \ref{fig:1}e to understand which evolution of the dispersing features is expected in the experiment. Taking into account the multiple hybridizations of the bands and spin-orbit interaction (especially strong in case of Te-containing material), we label the bands according to their energy at zero in-plane momentum, as shown in panel with $k_z$=0. The first counter-intuitive observation is that up to $k_z\sim$0.2 all four dispersions stay nearly intact in spite of strongly dispersing $p_z$-feature. Only v-shape deviations within small momentum interval near $k_x=k_y=0$ signal the presence of two Dirac points and inverted band gap. If such tiny deviations are to be detected by ARPES, very high energy and momentum resolutions are indispensable. With further increase of $k_z$, $p_z$-band becomes easily traceable by following the top of  feature \#1. It is in this way that its presence has been detected for the first time in LiFeAs \cite{borisenko2016direct} and if the band inversion occurs also in FeTe$_{0.55}$Se$_{0.45}$, this is the way to experimentally prove it.

\begin{figure*}
	\centering
		\includegraphics[width=0.95\linewidth]{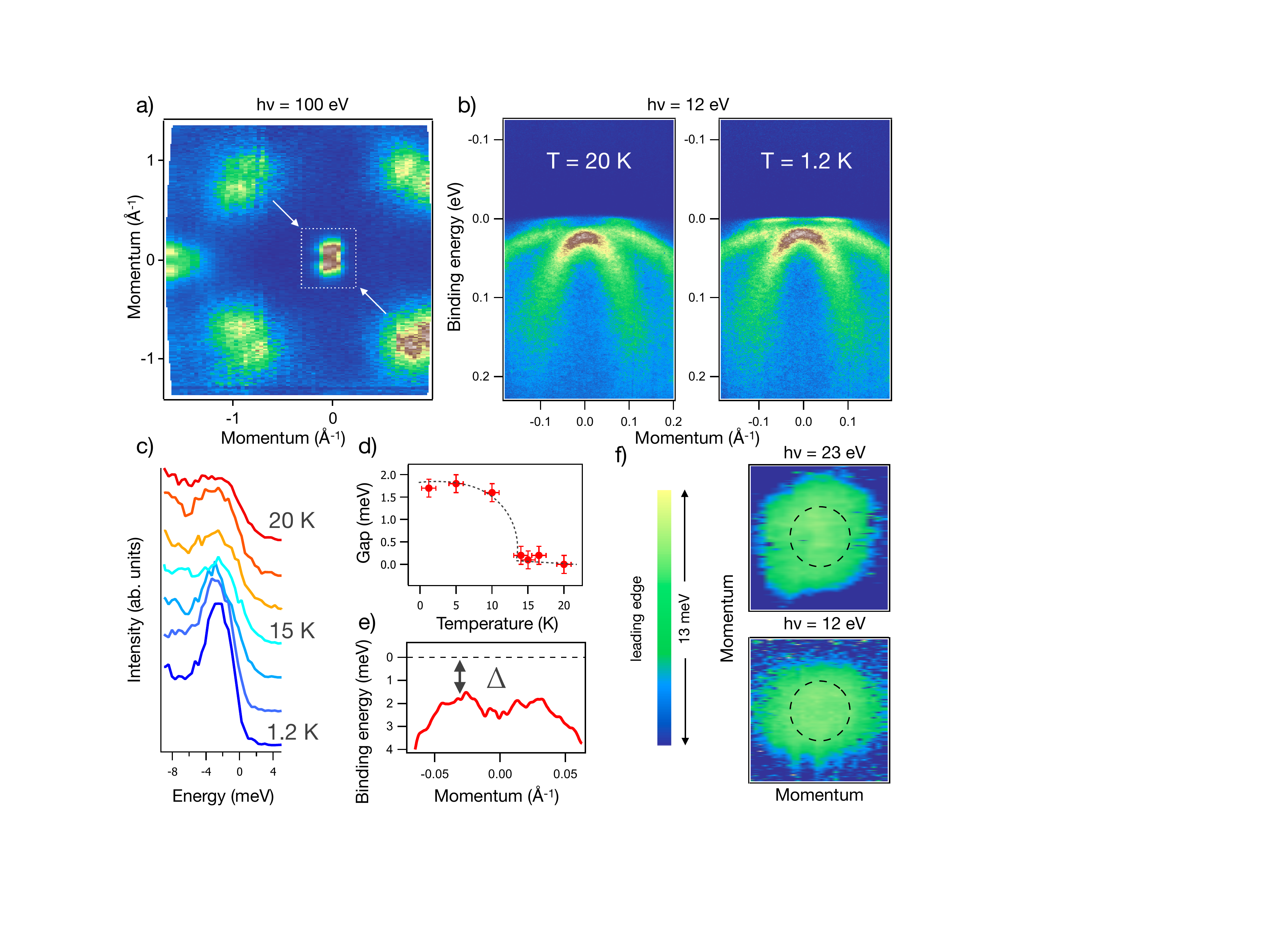}
	\caption{(a) Fermi surface map taken with 100 eV photons. (b) Data along the cut corresponding to white arrows in (a) at 20 and 1.2 K. (c) Temperature dependence of the EDC from $k_F$ in plots (b). (d) Superconducting energy gap as a function of temperature. (e) EDC dispersion from (b) at 1.2 K. (f) Maps of the leading edges of the EDCs from the area on the map (a) surrounded by the dashed line. }
	\label{fig:2}
\end{figure*}

We start presenting experimental data by showing the Fermi surface map recorded using 100 eV photons in Fig. \ref{fig:2}a. Apart from typical for IBS hole-like pocket in the middle and electron-like pockets in the corners of the BZ, such map demonstrates the excellent quality of the cleave - the intensity distribution from the large portion of the $k$-space is rather homogeneous and does not include any shadows from disoriented crystallites, common for FeSe \cite{maletz2014unusual, fedorov2016effect, Watson2015a}. Also, in line with other experiments \cite{fanfarillo2019photoinduced, okazaki2012evidence}, we do not notice any sign of nematic ordering on this map. Fig. \ref{fig:2}b compares photoemission intensity recorded along the diagonal cut (marked by white arrows in Fig. \ref{fig:2}a) at two temperatures: above and below T$_c$ of 13.5 K. One can notice an emergence of the superconducting coherence peak near the Fermi level for the relatively wide range of momentum values. Energy distribution curve (EDC) at Fermi momentum ($k_F\sim $0.03\AA$^{-1}$) clearly indicates the opening of the superconducting gap upon lowering the temperature (Fig. \ref{fig:2}c,d). Peak position of the EDC at 1.2 K as a function of momentum allows one to extract both the gap value and $k_F$ itself. We also present two maps of gaps (momentum distribution of the leading edge position within the area marked by dashed line in Fig. \ref{fig:2}a) at two photon energies in Fig. \ref{fig:2}f to illustrate rather isotropic character of the order parameter in FeTe$_{0.55}$Se$_{0.45}$. This is in contrast to pristine FeSe, where the anisotropy due to nematicity has been observed \cite{kushnirenko2018three, rhodes2018scaling}.

The datasets shown in Fig. \ref{fig:2}b correspond to $k_z$=0 case and indicate the presence of three dispersing features \#1, \#2 and \#3, as in Fig. \ref{fig:1}e, considering the different position of the Fermi level. However, proximity of the tops of bands \#1 and \#2 at zero momentum (as in Fig. \ref{fig:1}e for $k_z$=0.365, where tops of bands \#2 and \#3 are close) implies that in terms of energetics of the Fe$d$-orbital manifold, one is closer to the cases shown in Fig. \ref{fig:1} a and b, where the Dirac crossing of the type II occurs very close to the $\Gamma$ point. The same datasets also allow us to extract the size of the spin-orbit splitting and renormalization factors. Division by the Fermi function (not shown) gives the location of the top of band \#3 at -2 meV. The spin-orbit splitting is thus equal to 32 meV, which is very reasonable, since it must exceed the one in pure FeSe ($\sim$20meV, subtracting nematic splitting \cite{kushnirenko2017anomalous}) because of the presence of Te atoms. Renormalization factors for bands \#1, \#2 and \#3 are 3, 5 and $\sim$20 respectively.

Note, that there is neither sign of band \#4 nor of the surface states in Fig. \ref{fig:2}b. While the absence of the former is natural and expected because it remains unoccupied at $k_z$=0 in three out of four cases discussed in Fig. \ref{fig:1}, the absence of the latter in our data is already a contradiction with Ref.\cite{zhang2018observation}, since the surface states are expected to be visible independently of the photon energy. Intriguingly, the size of the superconducting gap and the way it opens in Fig. \ref{fig:2} c-f on the bulk band in our study are very similar to the ones observed earlier for the topological surface states \cite{zhang2018observation}.

\begin{figure*}
	\centering
	    \includegraphics[width=0.95\linewidth]{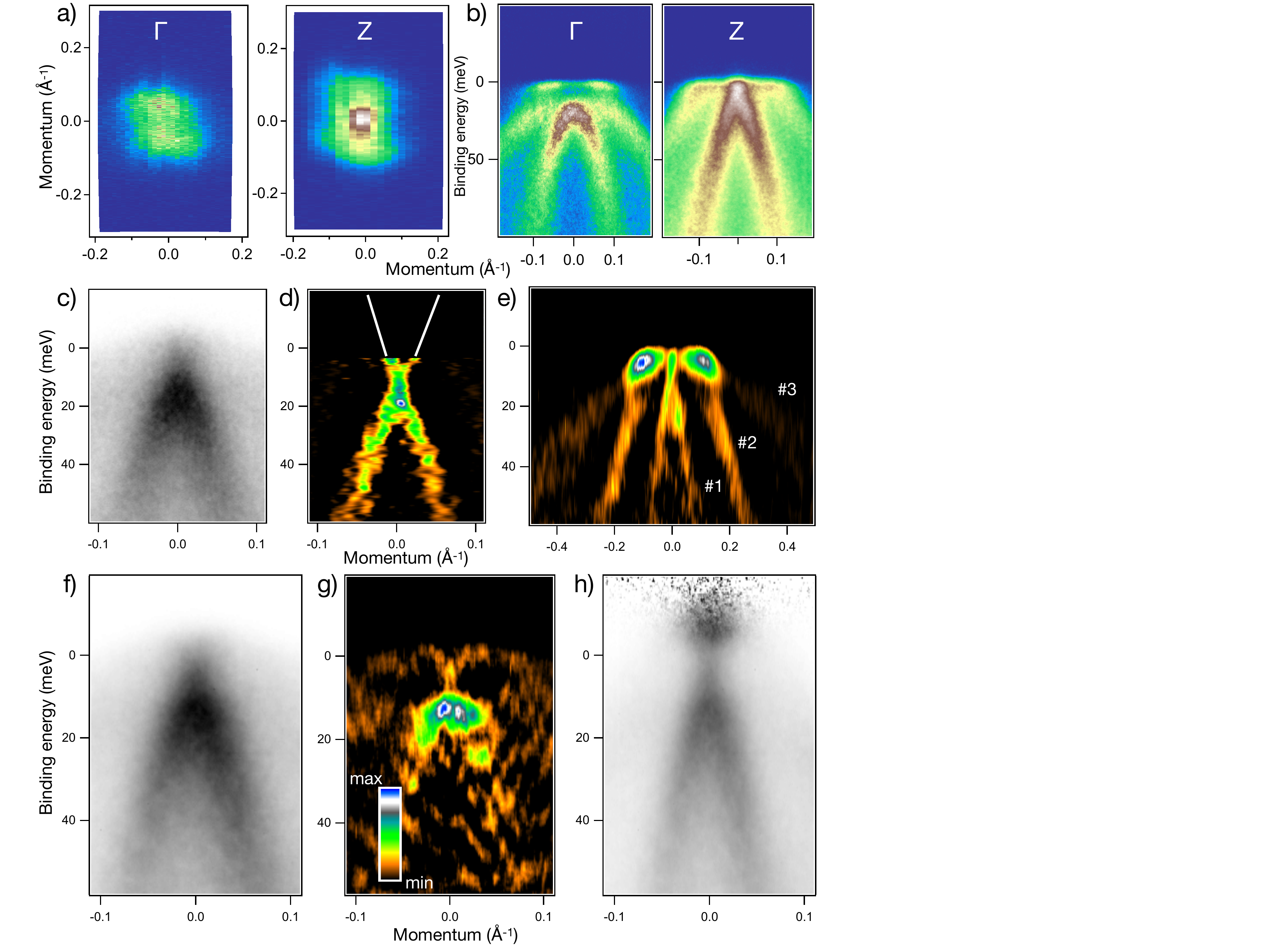}
	\caption{(a)Fermi surface maps recorded using 12 eV and 23 eV photon energies. (b) Diagonal cuts from the maps in (a). (c) Data taken with 18 eV photons at 20 K along the same diagonal direction in the BZ. (d) Second derivative of (c) divided by Fermi function. Noisy artefacts above the Fermi level are not shown. White lines - guides to the eye. (e) Second derivative in energy of the diagonal cut ($h\nu$=18 eV). (f-h) Diagonal cut with second derivative in energy and divided by Fermi function measured at 15 K and using 18 eV photons.}
	\label{fig:3}
\end{figure*}

In order to pursue the evolution of the electronic structure with $k_z$, we recorded the Fermi surface maps in the vicinity of $\Gamma$ and Z points using the 12 and 23 eV photons (Fig. \ref{fig:3}a). The contour at $\Gamma$ is smaller and does not contain the bright feature right in the center. Corresponding diagonal cuts in a narrow energy range very close to Fermi level are compared in Fig. \ref{fig:3}b. First, the qualitatively different spectra prove the strong photon energy sensitivity and thus imply a reasonable $k_z$-resolution, which is not always the case, even in IBS \cite{fedorov2019energy}. This is supported by the fact that not only the intensity distribution is different, but also the peaks of the spectral function obviously change their energy and momentum positions. Dispersion \#1 becomes sharper at zero momentum, mirroring the transformation seen in Fig. \ref{fig:1}e when going from $k_z$=0 to $k_z$=0.185. The bright spot on the map around Z (Fig. \ref{fig:3}a) is obviously caused by an additional feature, which appeared from above the Fermi level. In order to clarify its origin, we optimized the experimental conditions by selecting the particular photon energy (18 eV) and by raising the temperature to close the superconducting gap and increase the intensity of the states above the Fermi level. Such dataset is shown in Fig. \ref{fig:3}c. Now it is seen, that the feature has a v-shape electron-like dispersion strongly resembling the aforementioned v-shape portions of the dispersions in Fig. \ref{fig:3}e, which highlighted the proximity to the Dirac point.

To elucidate blurred features in ARPES spectra, the usual practice is to consider a second-derivative or curvature plots. However, this technique can be misguiding if applied to the closely separated dispersions in particular cases. Such example can be seen in Fig. 3d. Second derivative taken along the momentum axis makes an impression that there are two linear dispersions crossing at approximately 10 meV. However, knowing that in the region of interest our bands have extrema, it is more instructive to consider the second derivative along the energy axis. We show example of such treatment in Fig. 3e, where all three dispersions are traceable in a larger momentum region. Coming back to the case of $h\nu$=18 eV at 20 K temperature, we demonstrate in Fig. 3 f,g,h how additional tools, like second derivative along energy (Fig. 3g) and division by the Fermi function (Fig. 3h), allow one to learn more about the feature located at the Fermi level. Now it is seen, that there is actually no crossing of linear bands. Instead, the v-feature belongs to band \#3 which is still separated from band \#2 by something like 14 meV.

The examples in Fig.3 imply that a more detailed photon energy dependence is needed to unambiguously assign all features of the spectra to the calculated dispersions. In Fig. 4 we present such energy dependence for the second derivative plots in a broad kz-momentum range, covering a couple of BZs. These data immediately convey the striking and robust result - there are no features dispersing deeper than 30 meV binding energy at zero momentum in any of the presented spectra. This is in sharp contrast to LiFeAs \cite{borisenko2016direct}, where the $p_z$-states at Z-point were detected at binding energies higher than 200 meV. Moreover, the highest binding energy feature monotonically disperses towards the Fermi level from $\Gamma$ to Z. This behavior essentially excludes scenarios from Fig. 1 c and d.

Two further observations from Fig. 4: (i) the photon energy at which the v-shape feature at the Fermi level starts to be visible is 18 eV, implying (in accordance with the lineshape evolution from Fig. 1e) that the Dirac crossing of the first type already occured at closer to $\Gamma$ momentum values. It remains invisible because it is located above the Fermi level; (ii) this feature is almost absent near the second Z-point (42-43 eV), making it possible to track the position of the top of band \#2. 

In Fig. 5a we show the Fermi surface map in $k_x$-$k_z$ plane together with the second derivatives of this map and of momentum distribution at two binding energies of 30 and 45 meV. All contours clearly confirm our assignment of high symmetry points along $k_z$ in terms of both size variation and intensity distribution of the features. The Fermi surface is the smallest near the $\Gamma$-point and the largest when $k_z$ is maximal. The relative intensity of the features varies as a function of $k_z$ due to the well-known matrix element effects. Because of the strong mixture of orbital characters due to spin-orbit interaction and hybridization, it is not easy to explain all relative photoemission intensity variations as a function of polarization or photon energy in terms of usual parity analysis with respect to reaction plane. In addition, the phase change can significantly change the intensity of a particular feature at the BZ boundary. Finally, the photoemission cross-sections of Fe $3d$, Se $4p$ and Te $4p$ electrons are noticeably changing functions in the photon energy range studied here. Because of this, all argumentation in this study is based solely on the energy and momentum positions of the peaks in the spectral function and not on relative intensity variations of the photoemission signal.

In Fig. 5 b,c we attempt to extract quantitative information as regards the energy positions of the EDC peaks close to the Fermi level. This is a non-trivial task, since there are several features in this narrow energy region and their relative intensities are strongly varying. Fig. 5b is the second derivative (in energy) plot along $k_z$, corresponding to zero in-plane momentum. It sums up the results from Fig. 4, which is instructive to compare with Fig. 1 (a-d). In Fig. 5c we show the energy positions of the EDC peaks in the raw data. Both plots unambiguously rule out the scenario put forward recently \cite{wang2015topological, zhang2018observation}, where $p_z$-band disperses through all Fe $d$-bands. The only feature which has a dispersion minimum at Z-point along $k_z$ is the bottom of v-shape portion of band \#3. It starts to be clearly visible at $h\nu$ = 18 eV, but does not move lower than $\sim$9 meV at  photon energies of 23-24 eV. Because it nearly merges with the top of the band \#2 there, it is hard to figure out whether the band inversion occured or not. The second Z-point corresponding to $h\nu\sim$42 eV clarifies the situation. Now the v-shape feature is very weak and the position of the top of the band \#2 can be defined more precisely as 12-13 meV. This explains the different behavior of the EDCs' peaks near Z-points in Fig. 5c. Since now we can always track the positions of both band extrema, we can rule out also the scenario from Fig. 1b, i.e. $p_z$-states do not reach the lower $d_{xz/yz}$ dispersion and thus no band inversion occurs in FeTe$_{0.55}$Se$_{0.45}$. Another interference of the tops of the bands (first seen in Fig. 2b) results in peculiar behavior of the EDCs' peaks in the shaded area of Fig. 5c. In this region the dispersion seems to flatten out. This is because of the Dirac crossing of type-II between $d_{xz/yz}$ and $d_{xy}$ dispersions.

We sketch our scenario in Fig. 5 (d-h). At $k_z$=0 the assignment of the bands is straightforward and the positions of three band maxima can be defined from Fig. 2 b and c. It is important to realize that the highest intensity corresponds to band \#1, meaning that the lowest one in energy at $\Gamma$-point is a $d_{xz/yz}$-component (Fig. 5e). This situation rather quickly changes as the strongest peak disperses towards the Fermi level and starts to correspond to band \#2, signifying the Dirac crossing of the type II. The top of band \#1 becomes of $d_{xy}$-character and thus hardly visible in the spectra dispersing only slightly between $\Gamma$ and Z, as in the calculations (blue curve in Fig. 1a). However, its presence slightly influences the zeroth momentum EDC's peak position as is seen in Fig. 5c (shaded area). The next event in this scenario is the Dirac crossing of type-I above the Fermi level (Fig. 5f), which is not seen in the spectra, but which is unavoidable, since the top of band \#3 moves further above the Fermi level (note the missing intensity at zero energy and momentum in spectra taken with 13 and 14 eV photons in Fig. 4) and $p_z$-sates make itself evident at 18 eV as v-shaped feature (see the precursor of it as re-appearing intensity at zero energy and momentum in spectra taken with 15, 16 and 17 eV photons, Fig. 4). The v-shape feature appears in the spectra when $p_z$-character defines band \#3 after type-I Dirac crossing. This event is represented by Fig. 5g. Finally, the gap between the tops of band \#2 and band \#3 continues to close being centered at increasing binding energies towards Z-point and we do not find any evidence for leveling this gap off or going through minimum, and thus no evidence for the fact of band inversion. The latter may be still realized if the $p_z$-band is further shifted down by increased Te content, as Fig. 1c implies. However, in order to bring this gap to the Fermi level, one would need additional hole-doping. Our scenario implies that the Dirac point of type II is located at $\sim$20 meV binding energy; the Dirac point of type I - at $\sim$-10 meV; the lower component of $d_{xz/yz}$ disperses from 30 meV at $\Gamma$ to 12-13 meV at Z; $d_{xy}$-states remain at $\sim$20 meV; taking into account the minimum of $p_z$ band at $\sim$9meV at Z and the position of Dirac point, we estimate that its maximum is at $\sim$-25 meV at $\Gamma$.

Comparing these numbers to the results of the calculations from Fig. 1a, one can see that all mentioned bands are renormalized along $k_z$ by a factor of $\sim$10. Thus, the inverted gap, which according to calculations is equal to 5.6 meV in fully Te-substituted case, is expected to be smaller than 1 meV, meaning that detection of such gap and corresponding topological surface states is a non-trivial task.

FeTe$_{0.55}$Se$_{0.45}$ emerges as a highly correlated material even in comparison with other IBS. The renormalization of $p_z$-states in e.g. LiFeAs \cite{borisenko2016direct} is $\sim$ 3, reducing the size of the inverted gap to $\sim$6 meV. Fig. 5 (e-h) also illustrates the formation of the unique extended singularity close to the Fermi level, which is always of high importance for the mechanism of high-temperature superconductivity \cite{borisenko2013superconductivity}. Not only strong correlations, but also the hybridization between $d_{xz/yz}$ and $d_{xy}$ states (which are closer to the Fermi level than in pristine FeSe by something like 30 meV) and interaction with $p_z$-states form a very flat band \#3  (seen directly e.g. in Fig. 2b).

Apart from that, as was demonstrated above, FeTe$_{0.55}$Se$_{0.45}$ is a politypic 3D Dirac semimetal with an interesting mechanism of formation of Dirac points. From one side it resembles Cd$_3$As$_2$ where the Dirac points are also located along $\Gamma$Z \cite{borisenko2014experimental, neupane2014observation}, from the other, it reminds transition metal dichalcogenides, where the bulk Dirac cones are formed from a single-orbital manifold \cite{bahramy2018ubiquitous}.

In conclusion, we used synchrotron-based angle-resolved photoemission spectroscopy and band structure calculations to study the electronic structure of FeTe$_{1-x}$Se$_x$ (x=0.45) in its three-dimensional Brillouin zone. We found that the material is characterized by a trivial band structure, i.e. does not possess an inverted energy gap, but instead is a 3D Dirac semimetal with the Dirac points of both types located in an immediate vicinity of the Fermi level. The calculations imply that the lower content of Se may indeed result in the band inversion, but the size of the inverted gap hosting the topological surface states will remain negligible because of strong correlations. Unique interactions between the Fe $d$- and chalcogen $p$-states result in an extended singularity close to Fermi level created by a single flat band making it both a strongly correlated superconductor and a politypic 3D Dirac semimetal.

\textit{Acknowledgements} We are grateful to Andrey Chubukov and Amit Kanigel for the fruitful discussions. This work was supported under DFG grant 1912/7-1 and BMBF project UKRATOP. SA acknowledges the DFG via AS523/4-1, MS via STU 695/1-1.

\bibliography{literature}

\begin{figure*}
	\centering
		\includegraphics[width=0.95\linewidth]{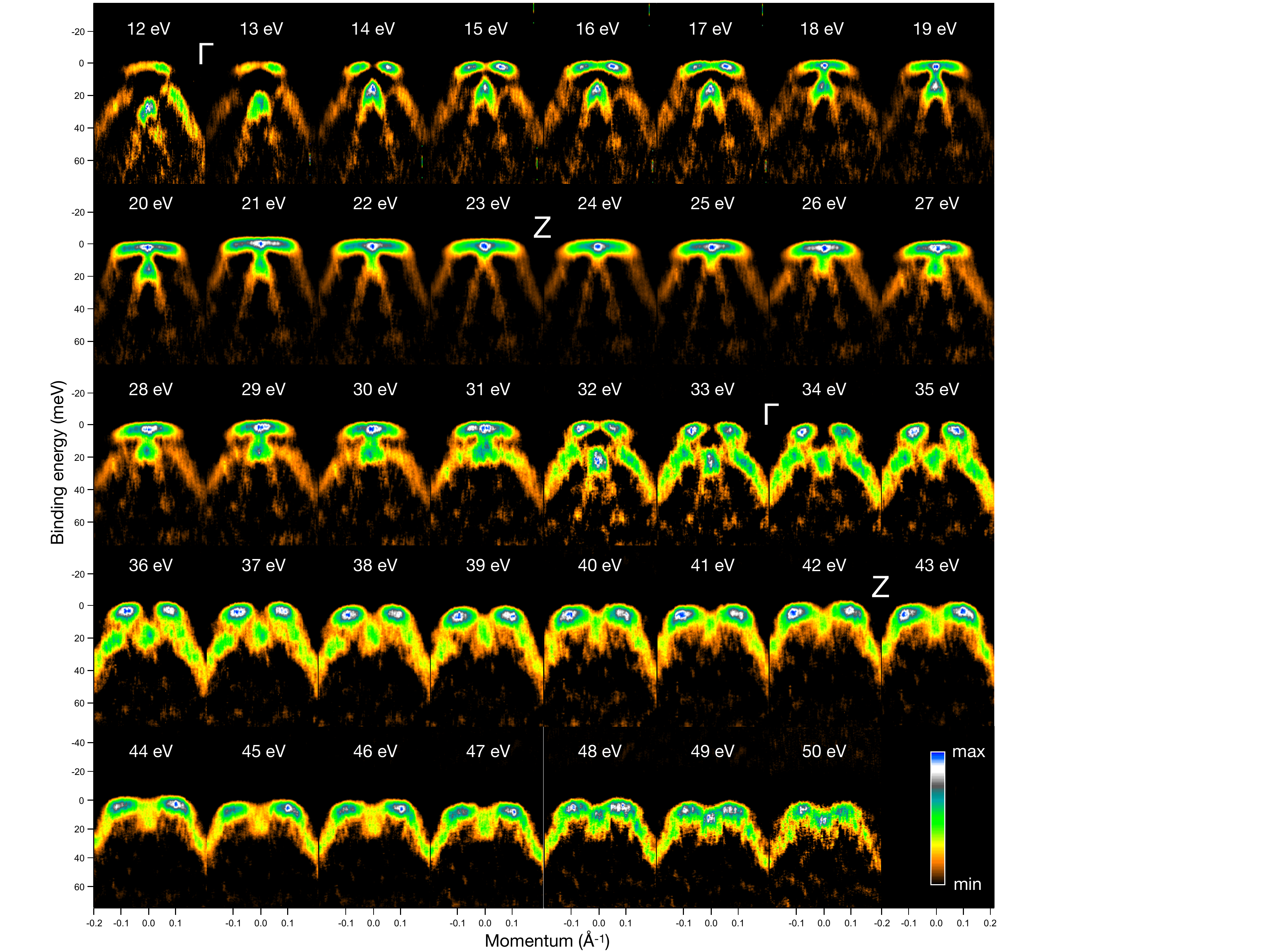}
	\caption{Photon energy dependence of the second derivative of the diagonal cut at 1 K.}
	\label{fig:4}
\end{figure*}

\begin{figure*}
	\centering
		\includegraphics[width=0.95\linewidth]{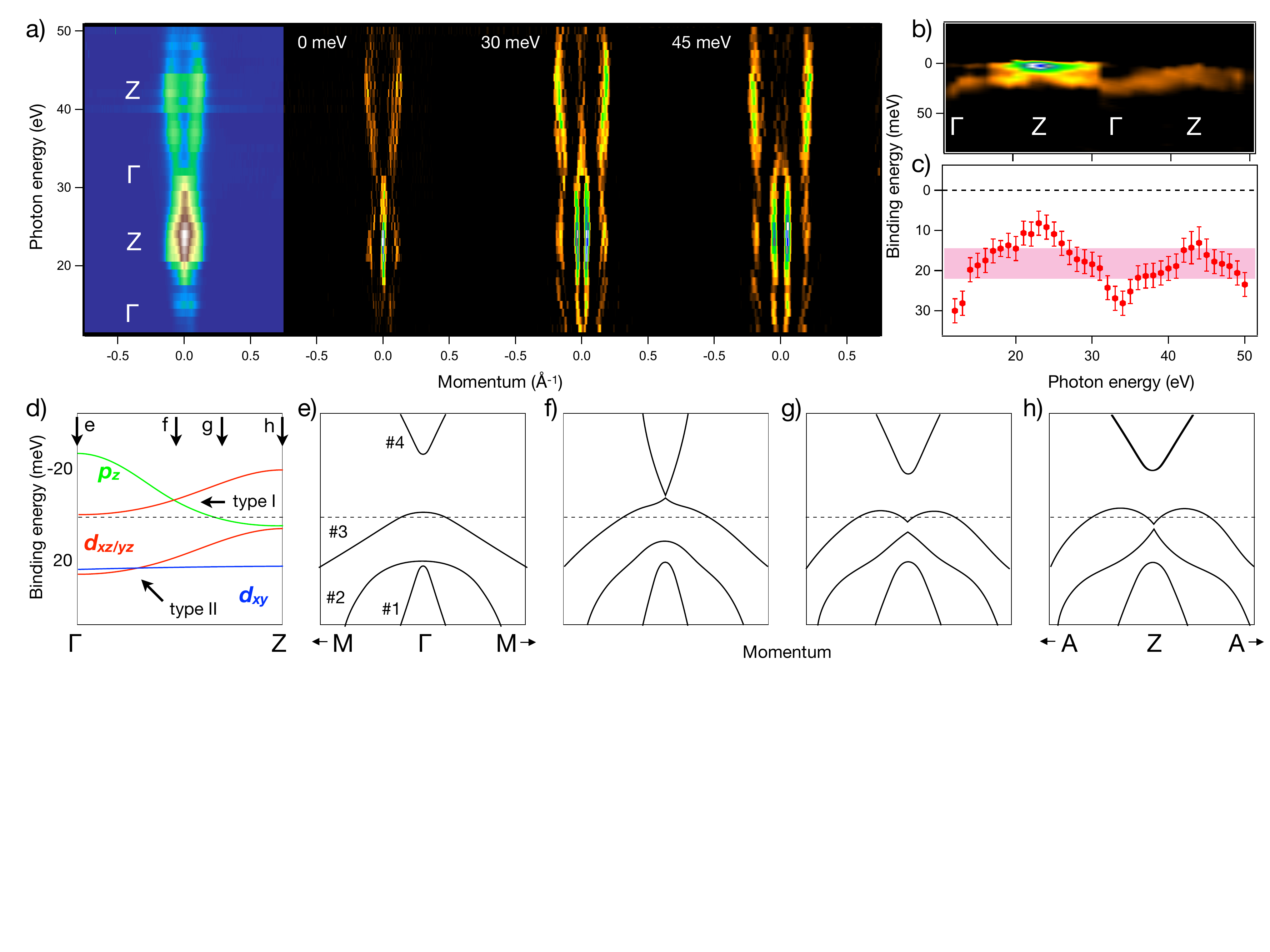}
	\caption{(a) Fermi surface map in $k_x-k_z$ plane and second derivatives of momentum distribution at 0, 30 and 45 meV binding energy. (b) Second derivative of the cut along $\Gamma$Z. (c) Peak positions of the EDC's maxima. (d-h) Schematic summary of the obtained results. }
	\label{fig:5}
\end{figure*}

\end{document}